\documentclass{iau-JDSS}
\usepackage{graphicx}
\title[Why coronal mass ejections are necessary for the dynamo]
{Why coronal mass ejections are necessary for the dynamo}

\author[Axel Brandenburg]   
{Axel Brandenburg}

\affiliation{Nordita, Blegdamsvej 17, DK-2100 Copenhagen \O, Denmark, and\\
AlbaNova University Center, SE - 106 91 Stockholm, Sweden
\break email: brandenb@nordita.dk}

\pubyear{2006}
\volume{Volume 14}  
\pagerange{1--2}
\date{?? and in revised form ??}
\jname{Highlights of Astronomy, Volume 14}
\editors{K.A. van der Hucht, ed.}
\begin{document}

\maketitle

\begin{abstract}
Large scale dynamo-generated fields are a combination of interlocked
poloidal and toroidal fields. Such fields possess magnetic helicity that
needs to be regenerated and destroyed during each cycle. A number of
numerical experiments now suggests that stars may do this by shedding
magnetic helicity. In addition to plain bulk motions, a favorite mechanism
involves magnetic helicity flux along lines of constant rotation. We also
know that the sun does shed the required amount of magnetic helicity
mostly in the form of coronal mass ejections. Solar-like stars without
cycles do not face such strong constraints imposed by magnetic helicity
evolution and may not display coronal activity to that same extent.
I discuss the evidence leading to this line of argument.
In particular, I discuss simulations showing the generation of strong
mean toroidal fields provided the outer boundary condition is left open
so as to allow magnetic helicity to escape. Control experiments with
closed boundaries do not produce strong mean fields.
\end{abstract}

\newcommand{\Fig}[1]{Fig.~\ref{#1}}
\newcommand{\bra}[1]{\langle #1\rangle}
\newcommand{\meanAA}{\overline{\mathbf{A}}}
\newcommand{\meanBB}{\overline{\mathbf{B}}}
\newcommand{\meanJJ}{\overline{\mathbf{J}}}
\newcommand{\meanUU}{\overline{\mathbf{U}}}
\newcommand{\meanWW}{\overline{\mathbf{W}}}
\newcommand{\meanSSSS}{\overline{\mathsf{S}}}
\newcommand{\meanFF}{\overline{{\cal{\mathbf{F}}}}}
\newcommand{\uu}{{\mathbf{u}}}
\newcommand{\BB}{{\mathbf{B}}}
\def\Rm{R_\mathrm{m}}

\begin{figure}\begin{center}
\includegraphics[width=.67\columnwidth]{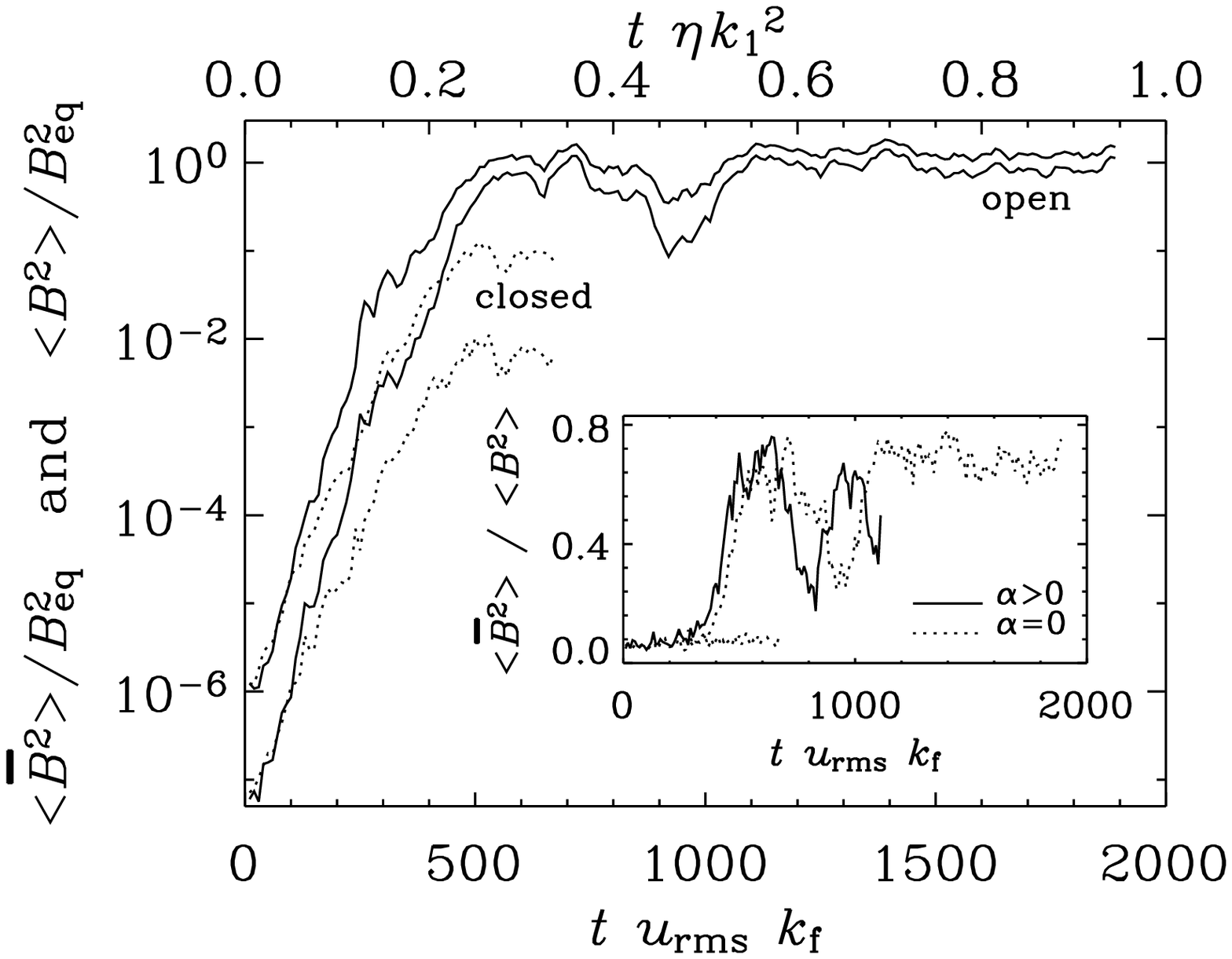}
\end{center}\caption[]{
Evolution of the energies of the total field $\bra{\BB^2}$ and of
the mean field $\bra{\meanBB^2}$, in units of $B_{\rm eq}^2$,
for runs with non-helical forcing
and open or closed boundaries; see the solid and dotted lines, respectively.
The inset shows a comparison of the ratio $\bra{\meanBB^2}/\bra{\BB^2}$
for nonhelical ($\alpha=0$) and helical ($\alpha>0$) runs.
For the nonhelical case the run with closed boundaries is also
shown (dotted line near $\bra{\meanBB^2}/\bra{\BB^2}\approx0.07$).
Adapted from Brandenburg (2005).
}\label{fig:pmean_comp}
\end{figure}

All known large scale dynamos ($\alpha\Omega$, shear--current,
and $\alpha^2$ dynamos) produce magnetic helicity, which reacts back on
the dynamo.
As a consequence, the mean field saturates at a low value,
$\meanBB^2\ll B_{\rm eq}^2\equiv\bra{\mu_0\rho\uu^2}$.
By allowing for magnetic helicity
fluxes out of the domain, the large scale field is able to saturate at
equipartition field strength (\Fig{fig:pmean_comp}).
The results of simulations are qualitatively, and in some cases also
quantitatively, well reproduced by mean field models where the effect of
magnetic helicity fluxes enters into the dynamical feedback formula for
the magnetic alpha effect (even when there is no kinetic alpha effect!).
For closed boundary conditions, the field saturates at much lower
strength and no large scale field is being produced.

\begin{figure}[t!]
\centering\includegraphics[width=.5\columnwidth]{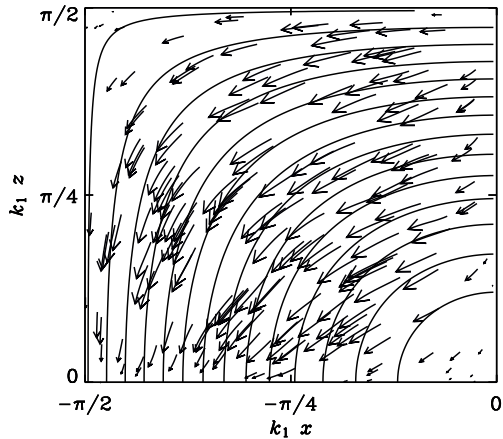}\caption{
Vectors of the Vishniac \& Cho flux together with contours of the mean
flow which also coincide with the streamlines of the mean vorticity field.
The orientation of the vectors indicates that negative current helicity
leaves the system at the outer surface at $x=0$.
The equator corresponds to $z=0$.
Note that the vectors indicate positive flux, so vectors pointing away
from the outer surface correspond to negative helicity leaving the sun
in the northern hemisphere.
Adapted from Brandenburg et al.\ (2005).
}\label{fig:p2vishcho}\end{figure}

Contributions to the magnetic helicity flux include the
shear-driven Vishniac-Cho (2001) flux, which can be
written in the form $\meanFF\propto(\meanSSSS\,\meanBB)\times\meanBB$,
where $\meanSSSS$ is the strain rate of the mean flow
(Subramanian \& Brandenburg 2006),
and an advectively driven flux (Shukurov et al.\ 2006) of the form
$\meanFF\propto\alpha_{\rm M}\meanUU$, where $\alpha_{\rm M}$ is the
magnetic $\alpha$ effect.
The former is the one operating predominantly in the simulations
(\Fig{fig:p2vishcho}).

A connection between dynamo-generated magnetic helicity fluxes
and coronal activity was first predicted by Blackman \& Field (2002).
There are at present no direct simulations of turbulent dynamos
that also include the relevant physics behind coronal mass ejections.
On the other hand, the observed magnetic helicity fluxes from coronal
mass ejections (Berger \& Ruzmaikin 2000) of around $10^{46}{\rm Mx}^2$
per cycle agrees with what is predicted from simulations
(Brandenburg \& Sandin 2004).

\newcommand{\yjgr}[3]{ #1, \textit{JGR} #2, #3.}
\newcommand{\yana}[3]{ #1, \textit{A\&A} #2, #3.}
\newcommand{\yapj}[3]{ #1, \textit{ApJ} #2, #3.}
\newcommand{\yan}[3]{ #1, \textit{AN} #2, #3.}
\newcommand{\ymn}[3]{ #1, \textit{MNRAS} #2, #3.}

\end{document}